\documentclass[12pt, a4paper]{article}

\usepackage{cite}
\usepackage{amsmath,amssymb}
\usepackage{comment}
\usepackage{multirow}
\usepackage[utf8]{inputenc}
\usepackage{bm}
\usepackage{accents}

\usepackage{ifpdf}
\ifpdf        
  \usepackage{graphicx, hyperref, xcolor}     
\else     
  \usepackage[dvipdfmx]{graphicx, hyperref, xcolor}     
 \fi
\usepackage{subcaption}
 
\definecolor{orangered}{HTML}{FF4500}
\definecolor{crimson}{HTML}{DC143C}
\definecolor{rossoferrari}{HTML}{D9073D}
\definecolor{steelblue}{HTML}{4682B4}
\definecolor{mediumblue}{HTML}{0000CD}
\definecolor{forestgreen}{HTML}{228B22}
\hypersetup{
setpagesize=false,
bookmarksnumbered=true,%
bookmarksopen=true,%
colorlinks=true,%
linkcolor=orangered,
urlcolor=steelblue,
citecolor=steelblue,
}

\usepackage[height=21.5cm,width=16.5cm,centering]{geometry}

\newlength{\dhatheight} 
\newcommand{\doublehat}[1]{%
    \settoheight{\dhatheight}{\ensuremath{\hat{#1}}}%
    \addtolength{\dhatheight}{-0.25ex}%
    \hat{\vphantom{\rule{1pt}{\dhatheight}}%
    \smash{\hat{#1}}}
}

\newcommand{\hyphen}{\,\mathchar`-\mathchar`-\,}

\begin{document}

\begin{titlepage}

\begin{center}

\hfill UT-19-09\\
\hfill DESY 19-069

\vskip .75in

{\Large \bf 
  Indirect Studies of 
  Electroweakly Interacting Particles\\ \vspace{2mm}
  at 100 TeV Hadron Colliders\\
}

\vskip .75in

{\large
Tomohiro Abe$^{(a,b)}$, So Chigusa$^{(c)}$, Yohei Ema$^{(d)}$, and Takeo Moroi$^{(c)}$
}

\vskip 0.25in

$^{(a)}${\em Institute for Advanced Research, Nagoya University,\\
Furo-cho Chikusa-ku, Nagoya, Aichi, 464-8602 Japan}\\[.3em]
$^{(b)}${\em Kobayashi-Maskawa Institute for the Origin of Particles and the Universe,\\
Nagoya University, Furo-cho Chikusa-ku, Nagoya, Aichi, 464-8602 Japan}\\[.3em]
$^{(c)}${\em Department of Physics, Faculty of Science,\\
The University of Tokyo, Bunkyo-ku, Tokyo 113-0033, Japan}\\[.3em]
$^{(d)}${\em DESY, Notkestra{\ss}e 85, D-22607 Hamburg, Germany}\\[.3em]

\end{center}
\vskip .5in

\begin{abstract}
  There are many extensions of the standard model that predict the
  existence of electroweakly interacting massive particles (EWIMPs),
  in particular in the context of the dark matter.  In this paper, we
  provide a way for indirectly studying EWIMPs through the precise
  study of the pair production processes of charged leptons or that of
  a charged lepton and a neutrino at future $100\,\mathrm{TeV}$
  collider experiments.  It is revealed that this search method is
  suitable in particular for Higgsino, providing us the $5\sigma$
  discovery reach of Higgsino in supersymmetric model with mass up to
  $850\,\mathrm{GeV}$.  We also discuss how accurately one can extract
  the mass, gauge charge, and spin of EWIMPs in our method.
\end{abstract}

\end{titlepage}


\renewcommand{\thepage}{\arabic{page}}
\setcounter{page}{1}
\renewcommand{\thefootnote}{$\natural$\arabic{footnote}}
\setcounter{footnote}{0}

\section{Introduction}

ElectroWeakly Interacting Massive Particles (EWIMPs) are theoretically
well-motivated particles that appear in many models beyond the
standard model (SM).  They are widely discussed in the context of the
dark matter (DM), with identifying an electrically neutral (or
milli-charged) component as the DM.  An attractive feature of this
scenario is that the vanilla thermal freeze-out scenario predicts the
correct amount of the relic abundance for the EWIMP mass range of
$\mathcal{O} (1\hyphen10)\,\mathrm{TeV}$, and this mass range is within the scope of
current and future experiments.  Well-known examples of the EWIMPs are Higgsino and
Wino that arise within the supersymmetric extension of the SM.
Assuming that Higgsino (Wino) is the lightest supersymmetric particle
and its stability is assured by the $R$-parity,
its thermal relic abundance becomes consistent with the DM abundance
if the mass is $1.1\,\mathrm{TeV}$~\cite{Cirelli:2007xd,
  ArkaniHamed:2006mb} ($2.9\,\mathrm{TeV}$~\cite{Hisano:2006nn,
  Moroi:2013sla, Beneke:2016ync, ArkaniHamed:2006mb}). 
Another example is the
minimal dark matter~\cite{Cirelli:2005uq, Cirelli:2007xd,
  Cirelli:2009uv}, where a particle with a large $SU(2)_L$ charge is
identified as the DM.  The stability is automatically assured since
operators that cause its decay are suppressed by the cut-off scale of
the theory thanks to the large $SU(2)_L$ charge, provided that one
chooses a correct combination of the charge and spin.  A $5$-plet
Majorana fermion with a mass of $\mathcal{O}(10)\,\mathrm{TeV}$ is the
most popular in this context, but there are also other possibilities,
including both scalar and fermionic particles.

EWIMPs are extensively searched for by many experiments, including DM
direct, indirect detections and collider searches (in particular, the
mono-$X$ search and the disappearing charged track search).  While
EWIMPs with relatively large $SU(2)_L$ charges such as Wino and the
5-plet fermion are promising for these searches, Higgsino is typically
more challenging to probe~\cite{Baer:2014cua}.
Given this situation, another search strategy is
proposed~\cite{Alves:2014cda, Gross:2016ioi,
Farina:2016rws,Harigaya:2015yaa, Matsumoto:2017vfu, Chigusa:2018vxz,
DiLuzio:2018jwd, Matsumoto:2018ioi} that probes EWIMPs via the
electroweak precision measurement at colliders.  It utilizes a pair
production of charged leptons or that of a charged lepton and a
neutrino, where EWIMPs affect the pair production processes through the
vacuum polarizations of the electroweak gauge bosons.  It is an indirect
search method in the sense that it does not produce on-shell EWIMPs as
final states. The current status and future prospects have been analyzed
for LHC, ILC, CLIC, and $100\,\mathrm{TeV}$
colliders~\cite{Mangano:2016jyj, Contino:2016spe, Golling:2016gvc},
indicating that it provides a promising way to probe Higgsino as well as
the other EWIMPs.  A virtue of this method is that it is robust against
the change of the lifetime and the decay modes of EWIMPs and whether an
EWIMP constitutes a sizable portion of the DM or not.  Another important
point is that, due to EWIMPs, the invariant mass distributions of the
final state particles show sharp dip-like behavior at the invariant mass
close to twice the EWIMP mass.  It helps us to distinguish the EWIMP
effects from backgrounds and systematic errors.

In this paper, we pursue this indirect search method further.  In
particular, we demonstrate that the indirect search method can be
applied not only to discover EWIMPs but also \textit{to investigate
  their properties, such as charges, masses, and spins.}  To be more
specific, in this paper we focus on the future prospect of the
indirect studies of EWIMPs at $100\,\mathrm{TeV}$ colliders such as
FCC-$hh$~\cite{Benedikt:2651300} and
SppC~\cite{CEPC-SPPCStudyGroup:2015csa, CEPC-SPPCStudyGroup:2015esa}.
We update our previous analysis~\cite{Chigusa:2018vxz}
that has considered only the neutral current (NC) processes (mediated
by photon and $Z$-boson) by including the charged current (CC)
processes (mediated by $W$-boson) as well, as in
Refs.~\cite{DiLuzio:2018jwd, Matsumoto:2018ioi}.  It is crucial not
only to improve the sensitivity but also to break some degeneracy
among different EWIMP charge assignments; the NC and CC processes
depend on different combinations of the $SU(2)_L$ and $U(1)_Y$
charges, and hence the inclusion of both processes allows us to
extract these charges separately.

The rest of this paper is organized as follows.  In
Sec.~\ref{sec:ewimp}, we discuss how EWIMPs affect the production
processes of a charged lepton pair and those of a charged lepton and a
neutrino.  There we see that the EWIMP correction to the cross section,
as a function of the lepton pair invariant mass, develops a dip-like
structure when the invariant mass is around twice the EWIMP mass.  This
feature is essential in distinguishing the EWIMP effect from backgrounds
and systematic errors, as discussed in detail in
Sec.~\ref{sec:analysis}.  Although we have to rely on the transverse
mass instead of the invariant mass for the CC process, a similar
dip-like structure appears in the transverse mass distribution.
Sec.~\ref{sec:analysis} is divided into three parts.  First, we explain
our fitting based statistical approach, in which we absorb various
sources of systematic errors into a choice of nuisance parameters.
Next, we study the result of the EWIMP detection reach, updating our
previous results~\cite{Chigusa:2018vxz} by taking into account the CC
processes.  We then move to our main focus of this paper, namely the
future prospect of the mass, charge, and spin determination of the
EWIMP.  Finally Sec.~\ref{seq:conclusion} is devoted to conclusions.

\section{EWIMP effect on the lepton production processes}
\label{sec:ewimp}

We investigate contributions of the EWIMPs to the Drell-Yan process
through the vacuum polarization of the electroweak gauge bosons at the
loop level. Throughout the paper, we assume that all the other beyond
the SM particles are heavy enough so that they do not affect the
following discussion.  After integrating out the EWIMPs, the effective
lagrangian is expressed as
\begin{align}
 \mathcal{L}_{\rm eff} = \mathcal{L}_{\rm SM} + C_2 g^2\, W_{\mu \nu}^a
 f\left(-\frac{D^2}{m^2}\right) W^{a\mu\nu} + C_1 g'^2\, B_{\mu\nu}
 f\left(-\frac{\partial^2}{m^2}\right) B^{\mu\nu},\label{eq_lag}
\end{align}
where $\mathcal{L}_{\rm SM}$ is the SM Lagrangian, $D$ is a covariant
derivative, $m$ is the EWIMP mass,\footnote{Here we neglect a small
  mass splitting among the $SU(2)_L$ multiplet.} $g$ and $g'$ are the
$SU(2)_L$ and $U(1)_Y$ gauge coupling constants, and $W_{\mu\nu}^a$
and $B_{\mu\nu}$ are the field strength associated with the $SU(2)_L$ and
$U(1)_Y$ gauge group, respectively.  The function $f(x)$ is defined as
\begin{align}
 f(x) = \begin{cases}
	 \displaystyle{\frac{1}{16\pi^2} \int_0^1 dy\, y(1-y) \ln (1 -
	 y(1-y)x - i0)} & {\rm (Fermion)},\\[5mm]
	 \displaystyle{\frac{1}{16\pi^2} \int_0^1 dy\, (1-2y)^2 \ln (1-
	 y(1-y)x - i0)} & {\rm (Scalar)},
	\end{cases}\label{eq_f}
\end{align}
where the first (second) line corresponds to a fermionic (scalar) EWIMP, respectively.
The coefficients $C_1$ and $C_2$ for an $SU(2)_L$ $n$-plet EWIMP with hypercharge $Y$ 
are given by
\begin{align}
 C_1 &= \frac{\kappa}{8} n Y^2,\label{eq_C1}\\
 C_2 &= \frac{\kappa}{8} I(n),\label{eq_C2}
\end{align}
where $\kappa = 1, 2, 8, 16$ for a real scalar, a complex scalar, a
Weyl or Majorana fermion, and a Dirac fermion, respectively.  The Dynkin index $I(n)$ for the $n$ dimensional representation of
$SU(2)_L$ is given by 
\begin{align}
  I(n) = \frac{1}{12} (n^3 - n),\label{eq_dynkin}
\end{align}
which is normalized so that $I(2) = 1/2$.  The coefficients are
uniquely determined by the representation of the EWIMPs. For example,
$(C_1, C_2) = (1, 1)$ for Higgsino, and $(C_1, C_2) = (0, 2)$ for
Wino.  We emphasize that, contrary to the usual effective field
theory, our prescription is equally applied when the typical scale of
the gauge boson four-momentum, $q$, is larger than the EWIMP mass
scale $m$ since we do not perform a derivative expansion of $f$ in
Eq.~\eqref{eq_lag}.  It is important because, as we see soon, the
effect of the EWIMPs are maximized when $\sqrt{q^2}\sim m$, where the
derivative expansion is not applicable.

\begin{table}[t]
  \centering
  \def\arraystretch{1.2}
  \begin{tabular}{c|cccccc}
    Fermion $f$ & $v_f^{(\gamma)}$ & $a_f^{(\gamma)}$ & $v_f^{(Z)}$ & $a_f^{(Z)}$ & $v_f^{(W)}$ & $a_f^{(W)}$ \\ \hline
    up-type quark & $\frac{2}{3}e$ & 0 & $(\frac{1}{4}-\frac{2}{3}s_W^2) g_Z$ & $-\frac{1}{4}g_Z$ & $\frac{1}{2\sqrt{2}}g$ & $-\frac{1}{2\sqrt{2}}g$ \\
    down-type quark & $-\frac{1}{3}e$ & 0 & $(-\frac{1}{4}+\frac{1}{3}s_W^2)g_Z$ & $\frac{1}{4}g_Z$ & $\frac{1}{2\sqrt{2}}g$ & $-\frac{1}{2\sqrt{2}}g$ \\
    lepton & $-e$ & 0 & $(-\frac{1}{4}+s_W^2)g_Z$ & $\frac{1}{4}g_Z$ & $\frac{1}{2\sqrt{2}}g$ & $-\frac{1}{2\sqrt{2}}g$ \\
  \end{tabular}
  \caption{Coefficients of the weak interaction defined as
    $\Gamma_f^{(V)} \equiv v_f^{(V)} + a_f^{(V)} \gamma_5$.  Here, $e = g
    s_W$ and $g_Z = g / c_W$, where $s_W \equiv \sin \theta_W$ and $c_W
    \equiv \cos \theta_W$ with $\theta_W$ being the weak mixing angle.}
  \label{table_weak}
\end{table}

At the leading order (LO), we are interested in $u(p)~\bar{u}(p') \to
\ell^{-}(k)~\ell^{+}(k')$ and $d(p)~\bar{d}(p') \to
\ell^{-}(k)~\ell^{+}(k')$ as the NC processes and $u(p)~\bar{d}(p')
\to \nu(k)~\ell^{+}(k')$ and $d(p)~\bar{u}(p') \to
\ell^{-}(k)~\bar{\nu}(k')$ as the CC processes.  Here, $u$ and $d$
collectively denote up-type and down-type quarks, respectively, and
$p, p', k$, and $k'$ are initial and final state momenta.  In the SM,
the amplitudes for both the NC and CC processes at the LO are
expressed as
\begin{align}
 \mathcal{M}_{\rm SM} = \sum_{V} \frac{\left[ \bar{v}(p')
 \gamma^\mu \Gamma_q^{(V)} u(p) \right] \left[ \bar{u}(k) \gamma_\mu
 \Gamma_{\ell}^{(V)} v(k') \right]}{s' - m_V^2},\label{eq_m_sm}
\end{align}
where $\sqrt{s'}$ is the invariant mass of the final state leptons,
which is denoted as $m_{\ell\ell}$ for the NC processes and
$m_{\ell\nu}$ for the CC processes.  The relevant gauge bosons are $V
= \gamma, Z$ for the NC processes and $V = W^\pm$ for the CC
processes, with $m_V$ being the corresponding gauge boson mass. In addition,
\begin{align}
  \Gamma_f^{(V)} \equiv v_f^{(V)} + a_f^{(V)} \gamma_5,
\end{align}
with $v_f^{(V)}$ and $a_f^{(V)}$ given in Tab.~\ref{table_weak}.
The EWIMP contribution is given by
\begin{align}
 \mathcal{M}_{\rm EWIMP} = \sum_{V,V'} C_{VV'} s' f\left(\frac{s'}{m^2}\right)
 \frac{\left[ \bar{v}(p') \gamma^\mu \Gamma_q^{(V)} u(p) \right]
 \left[ \bar{u}(k) \gamma_\mu \Gamma_\ell^{(V')} v(k') \right]}
 {(s'-m_V^2)(s'-m_{V'}^2)},\label{eq_m_ewimp}
\end{align}
where $C_{\gamma \gamma} = 4(C_1 g'^2 c_W^2 + C_2 g^2 s_W^2)$,
$C_{\gamma Z} = C_{Z \gamma} = 4(C_2 g^2 - C_1 g'^2) s_W c_W$, $C_{Z
  Z} = 4(C_1 g'^2 s_W^2 + C_2 g^2 c_W^2)$, and $C_{WW} = 4 C_2 g^2$.
Again $V, V' = \gamma, Z$ for the NC processes and $V, V' = W^\pm$ for
the CC processes.

We use $d\Pi_{\mathrm{LIPS}}$ for a Lorentz invariant phase space factor for the two particles final state.  Then, using Eqs.~\eqref{eq_m_sm} and~\eqref{eq_m_ewimp}, we define
\begin{align}
 \frac{d \sigma_{\rm SM}}{d\sqrt{s'}} &= \sum_{a,b}
 \frac{dL_{ab}}{d\sqrt{s'}} \int d\Pi_{\mathrm{LIPS}}\, \left| \mathcal{M}_{\rm SM} \left( q_a
 \bar{q}_b \to \ell\ell / \ell\nu \right) \right|^2,\label{eq_sig_sm}\\
 \frac{d \sigma_{\rm EWIMP}}{d\sqrt{s'}} &= \sum_{a,b}
 \frac{dL_{ab}}{d\sqrt{s'}} \int d\Pi_{\mathrm{LIPS}}\, 2 \Re \left[ \mathcal{M}_{\rm SM}
 \mathcal{M}_{\rm EWIMP}^{*} \left( q_a \bar{q}_b \to \ell\ell / \ell\nu
 \right) \right],\label{eq_sig_ewimp}
\end{align}
where we take the average and summation over spins.  Here,
$dL_{ab} / d\sqrt{s'}$ is the luminosity function for a fixed
$\sqrt{s'}$:
\begin{align}
 \frac{d L_{ab}}{d\sqrt{s'}} \equiv \frac{1}{s} \int_0^1 dx_1
 dx_2~f_a(x_1) f_b(x_2) \delta\left(\frac{s'}{s} - x_1 x_2\right),
\end{align}
where $a$ and $b$ denote species of initial partons, $\sqrt{s}$ is the
center of mass energy of the proton collision ($\sqrt{s} = 100\,{\rm
  TeV}$ in our case), and $f_a(x)$ is a parton distribution function
(PDF) of the given parton $a$.  Eq.\,\eqref{eq_sig_sm} represents the
SM cross section, while Eq.\,\eqref{eq_sig_ewimp} the EWIMP
contribution to the cross section.  For the statistical treatment in
the next section, we introduce a parameter $\mu$ that parametrizes the
strength of the EWMP effect, and express the cross section with $\mu$
as
\begin{align}
 \frac{d\tilde{\sigma}}{d\sqrt{s'}} = 
 \frac{d\sigma_{\rm SM}}{d\sqrt{s'}} 
 + \mu \frac{d\sigma_{\rm EWIMP}}{d\sqrt{s'}}.
 \label{eq_diffcrosssection}
\end{align}
Obviously, $\mu=0$ corresponds to the pure SM, while $\mu=1$ corresponds to
the SM$+$EWIMP model.  Hereafter, we use
\begin{align}
 \delta_\sigma ( \sqrt{s'} ) \equiv \frac{d\sigma_{\rm
 EWIMP} / d\sqrt{s'}}{d\sigma_{\rm SM} /
 d\sqrt{s'}},\label{eq_dsigma}
\end{align}
to denote the correction from the EWIMP.

\begin{figure}[t]
 \centering
 \includegraphics[width=0.5\hsize]{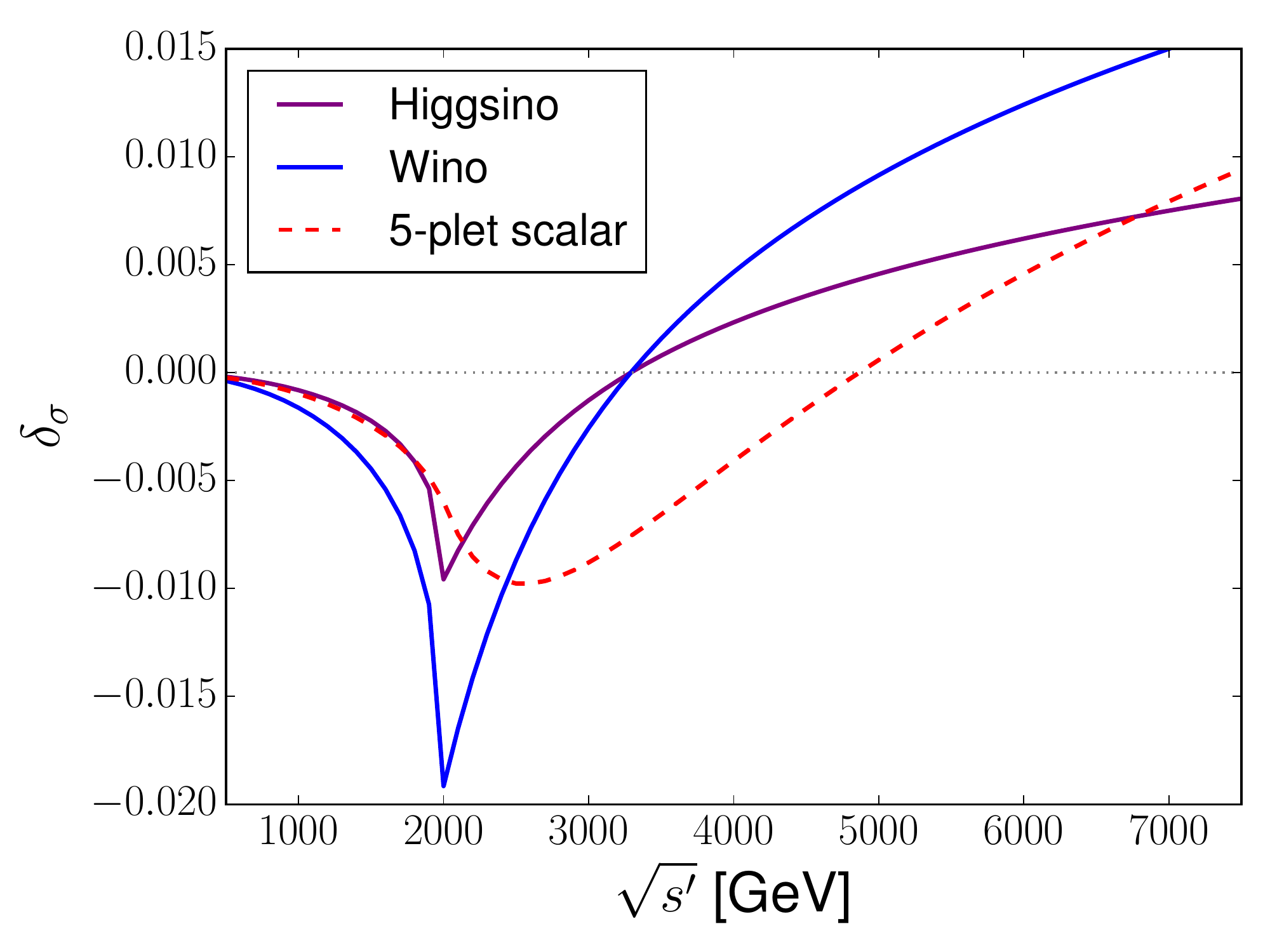}
 \caption{$\delta_\sigma$ for the CC processes as a
 function of $\sqrt{s'} = m_{\ell\nu}$.  The purple, blue, and red
 lines correspond to Higgsino, Wino, and 5-plet real scalar,
 respectively.}  \label{fig_sqsp_vs_del}
\end{figure}

In Fig.~\ref{fig_sqsp_vs_del}, we plot $\delta_\sigma$ for the CC
processes as a function of $\sqrt{s'}$.  The purple, blue, and red
lines correspond to Higgsino, Wino, and 5-plet scalar, respectively.
There is a dip around $\sqrt{s'} = 2m$ for all the cases of the EWIMPs
which originates from the loop function $f$ in Eq.~\eqref{eq_f}.  The
EWIMP contributions to the NC processes show a similar dip structure
that again comes from $f$.  This dip is crucial not only for the
discovery of the EWIMP signal (see Sec.~\ref{sec_detection}) but also
for the determination of the properties of the EWIMPs (see
Sec.~\ref{sec_property}).  In particular, the EWIMP mass can be
extracted from the dip position, while the EWIMP charges ($n$ and $Y$)
can be determined from the depth of the dip.

\begin{figure}[t]
 \centering
 \includegraphics[width=0.5\hsize]{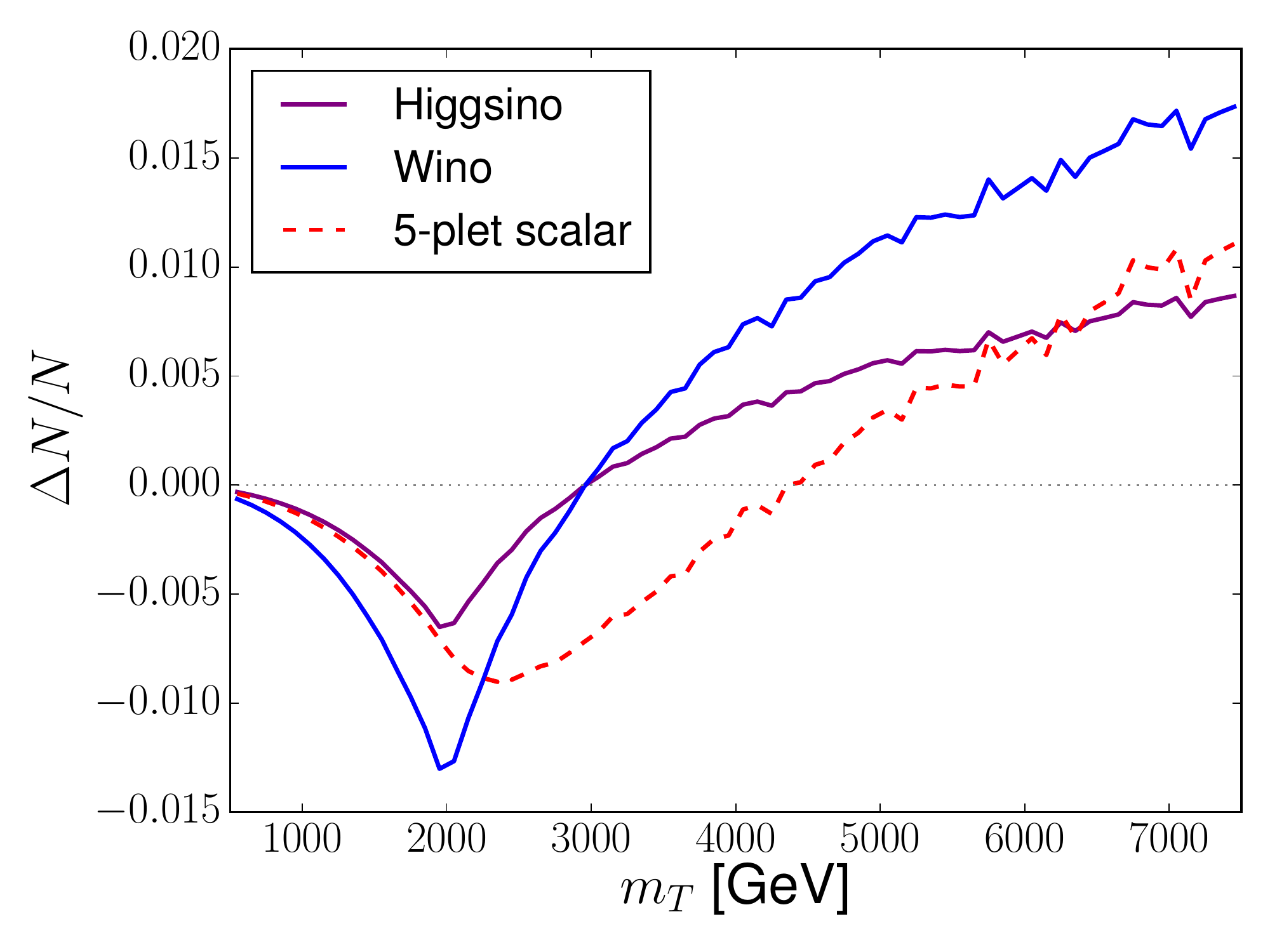}
 \caption{The EWIMP effect on the ratio of the number of events $\Delta
   N / N$ as a function of $m_T$.  The line colors are the same as
   Fig.~\ref{fig_sqsp_vs_del}.}  \label{fig_mT_vs_dN}
\end{figure}

For the NC processes, the momenta of two final state charged leptons are
measurable and we can use the invariant mass distribution of the number
of events for the study of the EWIMPs.  For the CC processes, on the
contrary, we cannot measure the momentum of the neutrino in real
experiments, and hence we instead use the missing transverse momentum
$p_{T,\mathrm{miss}}$.  We use the transverse mass defined as
\begin{align}
 m_T^2 \equiv 2 p_{T, \ell}\,p_{T,\mathrm{miss}} \left( 1-\cos
 (\phi_{T,\ell,\mathrm{miss}}) \right),
\end{align}
where $p_{T, \ell}$ denotes the transverse momentum of the charged
lepton and $\phi_{T,\ell,\mathrm{miss}} \equiv \phi_{\ell} -
\phi_\mathrm{miss}$ is the difference between the azimuth angles of
$p_{T,\ell}$ and $p_{T,\mathrm{miss}}$.  The important property of $m_T$
is that the distribution of $m_T$ peaks at $m_T = m_{\ell\nu}$.  Because
of this property, the characteristic shape of $\delta_\sigma$ remains in
the $m_T$ distribution in the CC events.  To see this, we plot in
Fig.~\ref{fig_mT_vs_dN} the EWIMP effect on the number of events as a
function of $m_T$.  Here, the vertical axis is the ratio of the EWIMP
correction to the number of events $\Delta N$ to the number of events in
the SM $N$ for each bin with the bin width of $100\, {\rm GeV}$.\footnote{
  Just for an illustrative purpose, we generate events corresponding
  to the integrated luminosity $\mathcal{L} = 1\,\mathrm{ab}^{-1}$ for
  this figure, which is not the same luminosity as we use in the next
  section (see Sec.~\ref{sec_event} for details of the event
  generation).
} We find that the dip structure remains in the $m_T$ distribution,
though the depth of the dip is smaller compared to the $m_{\ell\nu}$
distribution.

\section{Analysis}
\label{sec:analysis}

\subsection{Event generation}
\label{sec_event}

Now we discuss how well we can extract information about EWIMPs from the
invariant mass and transverse mass distributions for the processes of
our concern at future $100\,\mathrm{TeV}$ $pp$ collider experiments.  We
take into account the effects of the next-to-leading order QCD
corrections in the events as well as detector effects through
Monte-Carlo simulations.

In our analysis, we first generate the SM event sets for the NC
processes $pp\to e^{-}e^{+} / \mu^{-}\mu^{+}$ and for the CC processes
$pp\to e^{\pm}\nu_e / \mu^{\pm}\nu_\mu$.  We use
\texttt{MadGraph5\_aMC@NLO} (\texttt{v2.6.3.2})~\cite{Alwall:2011uj,
  Alwall:2014hca} for the event generation with the successive use of
\texttt{Pythia8}~\cite{Sjostrand:2014zea} for the parton shower and
the hadronization and \texttt{Delphes}
(\texttt{v3.4.1})~\cite{deFavereau:2013fsa} for the detector
simulation.  We use \texttt{NNPDF2.3QED} with $\alpha_s (M_Z) =
0.118$~\cite{Ball:2013hta} as a canonical set of PDFs.  For the
renormalization and factorization scales, we use the default values of
\texttt{MadGraph5\_aMC@NLO}, i.e., the central $m_T^2$ scale after
$k_T$-clustering of the event (which we denote by $Q$).  The events
are binned by the characteristic mass $m_{\mathrm{char}}$ for each
process: we use the lepton invariant mass $m_{\mathrm{char}} =
m_{\ell\ell}$ for the NC processes, and the transverse mass
$m_{\mathrm{char}} = m_T$ for the CC processes, respectively.  In both
cases, we generated events with the characteristic mass within the
range of $500\,\mathrm{GeV} < m_\mathrm{char} < 7.5\,\mathrm{TeV}$ and
divide them into $70$ bins with the equal width of
$100\,\mathrm{GeV}$.

As for the event selection by a trigger, we may have to impose some cut
on the lepton transverse momentum $p_T$.  As we will see, we concentrate
on events with high $p_T$ charged lepton(s) with which we expect the
event may be triggered.  For the NC processes, we use events with at
least two high $p_T$ leptons.  For our analysis, we use events with
$m_{\ell\ell}>500\ {\rm GeV}$; we assume that such events are triggered
by using two energetic charged leptons so that we do not impose extra
kinematical requirements.  On the contrary, the CC events are
characterized only by a lepton and a missing transverse momentum.  For
such events, we require that the $p_T$ of the charged lepton should be
larger than $500\,\mathrm{GeV}$.\footnote{
In the ATLAS analysis of the mono-lepton signal during the 2015 (2016)
data taking period~\cite{Aaboud:2017efa}, they use the event selection
condition $p_T > 24\, (60)\,\mathrm{GeV}$ for leptons that satisfy the
\textit{medium} identification criteria.  In the CMS analysis during
the period on 2016~\cite{Sirunyan:2018mpc}, they use the condition
$p_T > 130 (53)\, \mathrm{GeV}$ for an electron (a muon).
} For the CC events, the cut reduces the number of events in particular
for the bins with the low transverse mass $m_T \sim 500\, \mathrm{GeV}$,
and thus affects the sensitivity of the CC processes to relatively light
EWIMPs.  We will come back to this point later.

The EWIMP effect is incorporated by rescaling the
SM event by $\delta_\sigma$ defined in Eq.~\eqref{eq_dsigma}.  With the
parameter $\mu$ defined in Eq.~\eqref{eq_diffcrosssection}, the number
of events corresponding to the SM+EWIMP hypothesis in $i$-th bin,
characterized by $m_{i, \mathrm{min}} < m_{\mathrm{char}} < m_{i, \mathrm{max}}$, is
\begin{align}
  x_{f,i} (\mu) = \sum_{m_{i, \mathrm{min}} < m_{\mathrm{char}} < m_{i, \mathrm{max}}} 
  \left[
    1 + \mu \delta_\sigma (\sqrt{s'})
  \right],
  \label{eq_n_tot}
\end{align}
where the sum runs over all the events of the final state $f$ whose characteristic mass
$m_{\mathrm{char}}$ (after taking into account the detector effects)
falls into the bin.  Note that the true value of $\sqrt{s'}$ should be used for
each event for the computation of $\delta_\sigma$: 
we extract it from the hard process information.\footnote{
	The $p_T$ cut for the CC process does not affect this estimation
	since the EWIMP does not modify the angular distribution of the
	final lepton and neutrino for the CC process.
}

\subsection{Statistical treatment}
\label{sec_statistical}

We now explain the statistical method we will adopt in our analysis.
We collectively denote our theoretical model as $\bm{x}_f(\mu) = \{
x_{f,i} (\mu) \}$, where $x_{f,i}(\mu)$ is given by
Eq.~\eqref{eq_n_tot}.  We denote the experimental data set as
$\check{\bm{x}}_f$ that in principle is completely unrelated to our
theoretical model $\bm{x}_f(\mu)$.  Since we do not have an actual
experimental data set for $100~\,\mathrm{TeV}$ colliders for now,
however, we take $\check{\bm{x}}_f = \bm{x}_f(\mu = 1)$ (for some
fixed values of the EWIMP mass and charges) throughout our analysis,
assuming that the EWIMP does exist.  In particular, this choice tests
the SM-only hypothesis if we take our theoretical model as
$\bm{x}_f(\mu=0)$.

If the expectation values of $x_{f,i} (\mu)$ are precisely known, the
sensitivity to EWIMPs can be studied only with statistical errors.  In
reality, however, the computation of $x_{f,i} (\mu)$ suffers various
sources of uncertainties, which results in systematic errors in our
theoretical model.  The sources include errors in the integrated
luminosity, the beam energy, choices of the renormalization and the
factorization scales, choices of PDF, the pile-up effect, higher order
corrections to the cross section, and so on.  In order to deal with
these uncertainties, we introduce sets of free parameters $\bm{\theta}_f
= \{ \theta_{f,\alpha} \}$ (i.e. nuisance parameters) which absorb
(smooth) uncertainties of the number of events, and modify our
theoretical model as
\begin{align}
\tilde{x}_{f,i} (\bm{\theta}_f, \mu) \equiv x_{f,i} (\mu)
 f_{\mathrm{sys}, i}(\bm{\theta}_f),\label{eq_xtilde}
\end{align}
where $f_{\mathrm{sys}, i}(\bm{\theta}_f)$ is a function that satisfies
$f_{\mathrm{sys}, i}(\bm{0}) =1$.  We expect that, if the function
$f_{\mathrm{sys}, i}$ is properly chosen, the true distribution of the
number of events in the SM is given by $\tilde{\bm{x}}_f
(\bm{\theta}_f,0) = \{ \tilde{x}_{f,i} (0) f_{\mathrm{sys},
i}(\bm{\theta}_f)\}$ for some value of $\bm{\theta}_f$.  In our
analysis, we adopt the five parameters fitting function given
by~\cite{Aaltonen:2008dn}
\begin{align}
 f_{\mathrm{sys}, i} (\bm{\theta}_f) =
 e^{\theta_{f,1}} (1 + \theta_{f,2} p_i)
 p_i^{(\theta_{f,3} + \theta_{f,4} \ln p_i + \theta_{f,5} \ln^2 p_i)},
\end{align}
where $p_i = 2m_{i} / \sqrt{s}$ with $m_i$ being the central value of
the lepton invariant mass (transverse mass) of the $i$-th bin for the NC
(CC) processes.  As we will see, the major effects of systematic errors
can be absorbed into $\bm{\theta}_f$ with this fitting function.

In order to test the SM-only hypothesis, we
define the following test statistic~\cite{Cowan:2010js}:
\begin{align}
  q_0 \equiv -2 \sum_{f=\ell \ell, \ell \nu} \ln \frac
  {L(\check{\bm{x}}_f ; \doublehat{\bm{\theta}}_f, \mu=0)}
  {L(\check{\bm{x}}_f ; \hat{\bm{\theta}}_f, \hat{\mu})}.
  \label{eq_q0}
\end{align}
Here, $\doublehat{\bm{\theta}}_f$ and $\{ \hat{\bm{\theta}}_f, \hat{\mu}
\}$ are determined so that
$\prod_f L(\check{\bm{x}}_f ; \bm{\theta}_f, \mu=0)$ and
$\prod_f L(\check{\bm{x}}_f ; \bm{\theta}_f, \mu)$ are maximized, respectively.
The likelihood function is defined as
\begin{align}
 L(\check{\bm{x}}_f ; \bm{\theta}_f, \mu) &\equiv
 L_{\bm{\theta}_f} (\check{\bm{x}}_f ; \mu) L'(\bm{\theta}_f ; \bm{\sigma}_f),\label{eq_L}
\end{align}
where
\begin{align}
 L_{\bm{\theta}_f} (\check{\bm{x}}_f ; \mu) &\equiv
 \prod_{i} \exp \left[
 -\frac{(\check{x}_{f,i} - \tilde{x}_{f,i} (\bm{\theta}_f, \mu))^2}
 {2 \tilde{x}_{f,i} (\bm{\theta}_f, \mu)}
 \right],\label{eq_Ltheta}\\
 L'(\bm{\theta}_f ; \bm{\sigma}_f) &\equiv
 \prod_{\alpha} \exp \left[
 - \frac{\theta_{f,\alpha}^2}{2\sigma_{f,\alpha}^2}
 \right].\label{eq_Lprime}
\end{align}
The product in Eq.\,\eqref{eq_Ltheta} runs over all the bins, 
while the product in Eq.\,\eqref{eq_Lprime} runs
over all the free parameters we introduced.  For each $\theta_{f,
  \alpha}$, we define the ``standard deviation'' $\sigma_{f, \alpha}$,
which parametrizes the possible size of $\theta_{f, \alpha}$ within
the SM with the systematic errors.  
If the systematic errors are negligible compared with the statistical error,
we can take $\bm{\sigma}_f \to \bm{0}$.  
We identify $( q_0
)^{1/2} = 5$ $(1.96)$ as the detection reach at the $5\sigma$
($95\,\%$ C.L.) level, 
since $q_0$ asymptotically
obeys a chi-square distribution with the degree of freedom one.

In order to determine $\bm{\sigma}_f$, 
we consider the following sources of the systematic errors:
\begin{itemize}
 \item Luminosity ($\pm 5\,\%$ uncertainty is assumed),
 \item Renormalization scale ($2Q$ and $Q/2$, instead of $Q$),
 \item Factorization scale ($2Q$ and $Q/2$, instead of $Q$),
 \item PDF choice (We use $101$ variants of \texttt{NNPDF2.3QED} with
       $\alpha_s (M_Z) = 0.118$~\cite{Ball:2013hta} provided by
       \texttt{LHAPDF6}~\cite{Buckley:2014ana} with IDs ranging from
       $244600$ to $244700$).
\end{itemize}
The values of $\bm{\sigma}_f$ are determined as follows.  Let
$\bm{y}_f$ be the set of number of events in the SM for the final
state $f$ with the canonical choices of the parameters, and
$\bm{y}'_f$ be that with one of the sources of the systematic errors being
varied.  We minimize the chi-square function defined as
\begin{align}
 \chi^2_f &\equiv \sum_i \frac
 {\left( y_{f,i}' - \tilde{y}_{f,i} (\bm{\theta}_f) \right)^2}
 {\tilde{y}_{f,i} (\bm{\theta}_f)},
\end{align}
where
\begin{align}
 \tilde{y}_{f,i} (\bm{\theta}_f) &\equiv
 y_{f,i} f_{\mathrm{sys},i} (\bm{\theta}_f),
\end{align}
for each final state $f$, and determine the best-fit values of
$\bm{\theta}_f$ for each set of $\bm{y}'_f$.  We repeat this process for
different sets of $\bm{y}'_f$, and $\bm{\sigma}_f$ are determined from
the distributions of the best-fit values of $\bm{\theta}_f$.  For
example, let us denote the best-fit values for the fit associated with
the luminosity errors $\pm 5\%$ as $\bm{\theta}_f^{\pm}$.
We estimate $\bm{\sigma}_f$ associated with these errors, denoted here as
$\bm{\sigma}_f^{\mathrm{lumi.}}$, as
\begin{align}
 \sigma_{f,\alpha}^{\mathrm{lumi.}} = \sqrt{\frac{(\theta_{f,\alpha}^{+})^2 + (\theta_{f,\alpha}^{-})^2}{N}},
\end{align}
where $N$ denotes the number of fitting procedures we have performed:
$N=2$ for this case.  
We estimate $\bm{\sigma}_f$ associated with
the other sources of the errors, denoted as
$\bm{\sigma}_f^{\mathrm{ren.}}$, $\bm{\sigma}_f^{\mathrm{fac.}}$, and
$\bm{\sigma}_f^{\mathrm{PDF}}$, in a similar manner.  
Finally, the total values of $\bm{\sigma}_f$ are obtained by combining all the
sources together as
\begin{align}
    \sigma_{f,\alpha} = \sqrt{(\sigma_{f,\alpha}^{\mathrm{lumi.}})^2
    + (\sigma_{f,\alpha}^{\mathrm{ren.}})^2
    + (\sigma_{f,\alpha}^{\mathrm{fac.}})^2
    + (\sigma_{f,\alpha}^{\mathrm{PDF}})^2}.
    \label{eq_comb_sig}
\end{align}

\begin{table}[t]
  \centering
  \begin{tabular}{c|ccccc}
  Sources of systematic errors & $\sigma_{ee,1}$ & $\sigma_{ee,2}$ & $\sigma_{ee,3}$ & $\sigma_{ee,4}$ & $\sigma_{ee,5}$ \\ \hline
  Luminosity: $\pm 5\,\%$ ($\bm{\sigma}_{ee}^{\mathrm{lumi.}}$) & $0.05$ & $0$ & $0$ & $0$ & $0$ \\
  Renormalization scale: $2Q, Q/2$ ($\bm{\sigma}_{ee}^{\mathrm{ren.}}$) & $0.4$ & $0.6$ & $0.3$ & $0.05$ & $0.004$ \\
  Factorization scale: $2Q, Q/2$ ($\bm{\sigma}_{ee}^{\mathrm{fac.}}$) & $0.3$ & $0.5$ & $0.2$ & $0.06$ & $0.004$ \\
  PDF choice ($\bm{\sigma}_{ee}^{\mathrm{PDF}}$) & $0.4$ & $0.7$ & $0.3$ & $0.06$ & $0.004$
  \end{tabular}
  \caption{Values of $\bm{\sigma}_{ee}$ for each source of systematic errors.  The result is the same for the $\mu\mu$ final state.}
  \label{tab_sys_ee}
\end{table}

\begin{table}[t]
  \centering
  \begin{tabular}{c|ccccc}
  Sources of systematic errors & $\sigma_{e \nu_e,1}$ & $\sigma_{e \nu_e,2}$ & $\sigma_{e \nu_e,3}$ & $\sigma_{e \nu_e,4}$ & $\sigma_{e \nu_e,5}$ \\ \hline
  Luminosity: $\pm 5\,\%$ ($\bm{\sigma}_{e \nu_e}^{\mathrm{lumi.}}$) & $0.05$ & $0$ & $0$ & $0$ & $0$ \\
  Renormalization scale: $2Q, Q/2$ ($\bm{\sigma}_{e \nu_e}^{\mathrm{ren.}}$) & $0.3$ & $0.4$ & $0.2$ & $0.04$ & $0.003$ \\
  Factorization scale: $2Q, Q/2$ ($\bm{\sigma}_{e \nu_e}^{\mathrm{fac.}}$) & $1.0$ & $1.6$ & $0.6$ & $0.1$ & $0.01$ \\
  PDF choice ($\bm{\sigma}_{e \nu_e}^{\mathrm{PDF}}$) & $0.6$ & $0.9$ & $0.4$ & $0.08$ & $0.006$
  \end{tabular}
  \caption{Best fit values of fit parameters for several sources of systematic errors for the $e\nu_e$ final state.  The result is the same for the $\mu\nu_\mu$ final state.}
  \label{tab_sys_ev}
\end{table}

\begin{table}[t]
 \centering
 \begin{tabular}{c|ccccc}
  Final state $f$ & $\sigma_{f,1}$ & $\sigma_{f,2}$ & $\sigma_{f,3}$ & $\sigma_{f,4}$ & $\sigma_{f,5}$ \\ \hline
  $ee$ & $0.7$ & $1.0$ & $0.4$ & $0.09$ & $0.008$ \\
  $\mu\mu$ & $0.7$ & $1.0$ & $0.4$ & $0.09$ & $0.008$ \\
  $e\nu_e$ & $1.2$ & $1.9$ & $0.7$ & $0.2$ & $0.01$ \\
  $\mu\nu_\mu$ & $1.2$ & $1.9$ & $0.7$ & $0.2$ & $0.01$ \\
 \end{tabular}
 \caption{Summary of standard deviations $\bm{\sigma}_f$ for each final
 state.}
 \label{tab_sys}
\end{table}

In Table~\ref{tab_sys_ee} and \ref{tab_sys_ev}, we show the values of
$\bm{\sigma}_{ee}$ and $\bm{\sigma}_{e\nu_e}$ associated with each
source of the systematic errors, respectively.  These values can be
interpreted as the possible size of the fit parameters within the SM,
which is caused by the systematic uncertainties.  As
explained in Eq.~\eqref{eq_comb_sig}, we combine these values in each
column to obtain $\bm{\sigma}_f$.  In Table~\ref{tab_sys}, we summarize
the result of the combination for all the final states.  
The values of $\bm{\sigma}_f$ are independent of the
final state lepton flavors since the energy scale of our concern is much higher than the lepton masses. 
However, we use different sets of
fit parameters $\bm{\theta}_{ee}$ and $\bm{\theta}_{\mu\mu}$ for the NC
processes and $\bm{\theta}_{e\nu_e}$ and $\bm{\theta}_{\mu\nu_\mu}$ for
the CC processes
because of the different detector response to electrons and muons.

\subsection{Detection reach}
\label{sec_detection}

\begin{figure}[t]
  \centering \includegraphics[width=0.7\hsize]{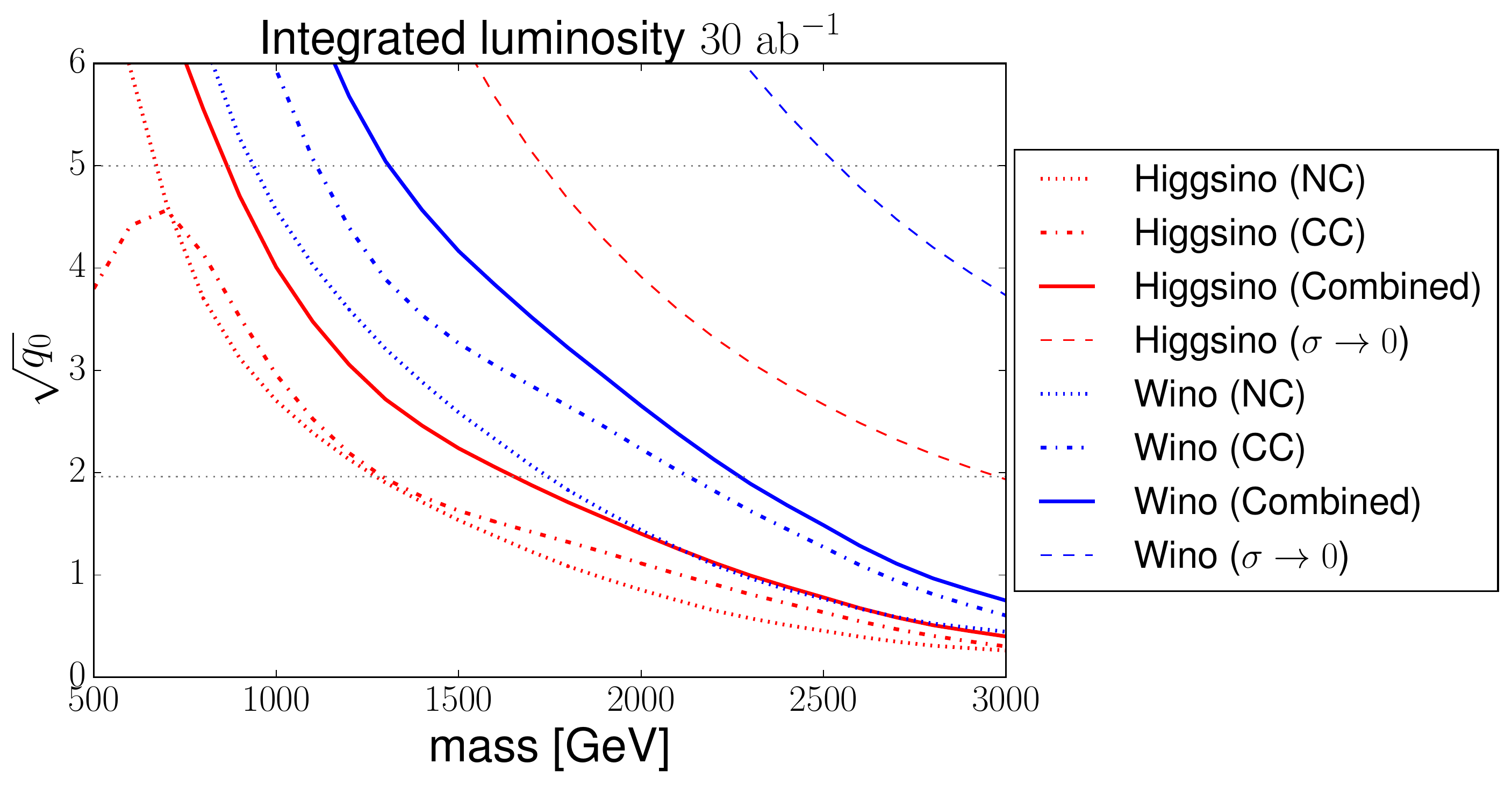}
  \caption{$\sqrt{q_0}$ as a function of the EWIMP mass.  Red and blue
    lines correspond to the Higgsino and the Wino, respectively, while
    line styles represent the result from the NC processes, the CC
    processes, the combined analysis, and the combined analysis with
    the optimistic $\bm{\sigma}_f \to \bm{0}$ limit.}
  \label{fig_mchi_vs_sqq0}
\end{figure}

Now we show the detection reach of EWIMPs at future
$100\,\mathrm{TeV}$ colliders.  In Fig.~\ref{fig_mchi_vs_sqq0}, we
plot the value of $\sqrt{q_0}$ as a function of the EWIMP mass, with
the integrated luminosity $\mathcal{L}=30\,\mathrm{ab}^{-1}$.  As
representative scenarios, we show the cases for Higgsino (the red
lines) and Wino (the blue lines).  The dotted and dash-dotted lines
are the result obtained only from the NC processes and the CC processes,
respectively.  We find that the CC processes are more sensitive to the
effect of the EWIMPs than the NC processes because of the larger cross
section.  This result is consistent with Refs.~\cite{DiLuzio:2018jwd,Matsumoto:2018ioi}.  
The sensitivity of the CC processes is weakened for $m \lesssim 700\,
\mathrm{GeV}$ 
because of the lepton $p_T$ cut we have applied.\footnote{
  We note here that the sensitivity of the CC processes depends on the
  lepton $p_T$ cut.  For example, adopting the tighter cut,
  lepton-$p_T > 1\,\mathrm{TeV}$, the CC processes have almost no
  sensitivity to EWIMPs with $m < 1\,\mathrm{TeV}$.  Thus, in
  particular for the purpose of the Higgsino search, it is important
  to realize the lepton $p_T$ cut as low as $\sim 500\, \mathrm{GeV}$.
} The combined results of the NC and CC processes are shown by the solid
lines.  By combining the two types of processes, the $5\sigma$ discovery
reaches ($95\,\%$ C.L. bounds) for Higgsino and Wino are
$850\,\mathrm{GeV}$ ($1.7\,\mathrm{TeV}$) and $1.3\,\mathrm{TeV}$
($2.3\,\mathrm{TeV}$), respectively.  
We find that the combination of
the NC and CC processes improves the sensitivity of the EWIMP mass. 
Furthermore, if we understand all the systematic uncertainties quite
well and effectively take the $\bm{\sigma}_f \to \bm{0}$ limit in the
combined result, the detection reach will be pushed up significantly as
shown by the dashed lines: $1.1\,\mathrm{TeV}$ Higgsino signal at well
above $5\sigma$ level and a $4\sigma$ hint of the $2.9\,\mathrm{TeV}$ Wino.
Therefore, it is essential to reduce the systematic uncertainties for
the detection of EWIMPs through the NC and CC processes.

\subsection{Determination of EWIMP properties}
\label{sec_property}

In this subsection, we show that it is possible to determine the
properties of the EWIMPs from the NC and CC processes, thanks to the
fact that we can study the $m_{\ell\ell}$ and $m_T$ distribution in
great detail for these processes.  Some information about the mass,
charge, and spin of the EWIMPs can be extracted because the
corrections to these distributions from the EWIMPs are completely
determined by these EWIMP properties.  Firstly, we can extract the
EWIMP mass from the position of the dip-like structure in the
correction since it corresponds to roughly twice the EWIMP mass as we
have shown in Sec.~\ref{sec:ewimp}.  Secondly, the overall size of the
correction gives us information about the $SU(2)_L$ and $U(1)_Y$
charges.  The CC processes depend only on the $SU(2)_L$ charge, while
the NC processes depend both on the $SU(2)_L$ and $U(1)_Y$ charges.
Consequently, we can obtain information about the gauge charges of the
EWIMPs from the NC and CC processes. 

We now demonstrate the mass and charge determination of fermionic
EWIMPs. This is equivalent to the determination of the parameter set
$(m, C_1, C_2)$.  We generate the data assuming the SM $+$ EWIMP model
($\mu=1$) with some specific values of $m, n, Y$, and $\kappa$, with
which we obtain $(m, C_1, C_2)$.  We fix $\mu = 1$ for our theoretical
model as well, and hence the theoretical predictions of the number of
events also depend on these three parameters, $\bm{x}_f = \bm{x}_f
(m, C_1, C_2)$.  We define the likelihood function $L(\check{\bm{x}}_f
; \bm{\theta}_f, m, C_1, C_2)$ in the same form as
Eqs.~\eqref{eq_xtilde} and~\eqref{eq_L} with the theoretical
prediction $\bm{x}_f$ now understood as a function of $(m, C_1, C_2)$,
not of $\mu$.\footnote
{As shown in Eqs.\ \eqref{eq_C1} and \eqref{eq_C2}, $C_1$ and $C_2$
  are positive quantities (and $C_2$ is discrete).  In the figures,
  however, we extend the $C_1$ and $C_2$ axes down to negative
  regions just for presentation purposes.}
The test statistic is defined as
\begin{align}
  q (m, C_1, C_2)
  \equiv -2 \sum_f \ln \frac
  {L(\check{\bm{x}}_f ; \doublehat{\bm{\theta}}_f, m, C_1, C_2)}
  {L(\check{\bm{x}}_f ; \hat{\bm{\theta}}_f, \hat{m}, \hat{C}_1, \hat{C}_2)},
  \label{eq_q0f}
\end{align}
where the parameters $(\{\hat{\bm{\theta}}_f\}, \hat{m}, \hat{C}_1,
\hat{C}_2)$ maximize $\prod_f L(\check{\bm{x}}_f ;
\bm{\theta}_f, m, C_1, C_2)$, while
$\doublehat{\bm{\theta}}_f$ maximize $L(\check{\bm{x}}_f ;
\bm{\theta}_f, {m}, {C}_1, C_2)$ for fixed values of $(m,
C_1, C_2)$.  It follows the chi-squared distribution with three
degrees of freedom in the limit of a large number of
events~\cite{Tanabashi:2018oca}.  The test statistic defined in this
way examines the compatibility of a given EWIMP model (i.e. a
parameter set $(m, C_1, C_2)$) with the observed signal.

\begin{figure}[t]
 \centering
 \includegraphics[width=0.5\linewidth]{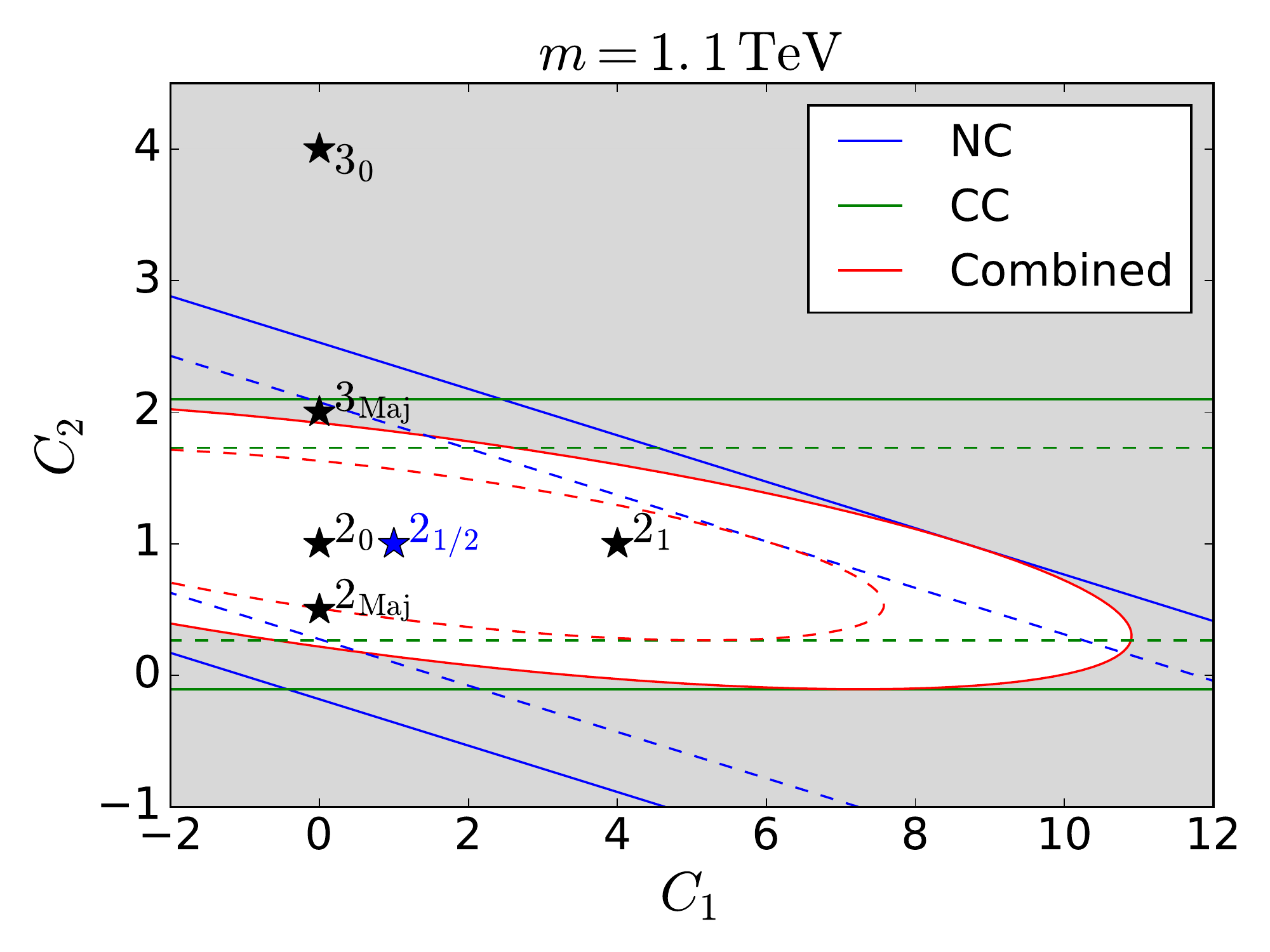}
 \caption{Contour of $\sqrt{q}$ in the $C_1~\mathrm{vs.}~C_2$ plane
 with $m = 1.1\,\mathrm{TeV}$,
 where we assume $1.1\,\mathrm{TeV}$ Higgsino signal. 
 The dotted and solid
 lines denote $1\sigma$ and $2\sigma$ contours, respectively, and the
 gray region corresponds to the parameter space that is in tension
 with the observation at more than $2\sigma$ level.  The blue, green,
 and red lines correspond to the result from the NC processes, the CC
 processes, and the combined analysis, respectively.  Each star marker
 annotated as ``$n_Y$'' represents a point corresponding to a
 $SU(2)_L$ $n$-plet Dirac fermion with hypercharge $Y$, while that
 with ``$n_\mathrm{Maj}$'' corresponds to an $SU(2)_L$ $n$-plet
 Majorana fermion.}
\label{fig_c1_c2}
\end{figure}

Once a deviation from the SM prediction is observed in a real experiment,
we may determine $(m, C_1, C_2)$ using the above test statistic $q$.
In the following, we
show the expected accuracy of the determination of $(m, C_1, C_2)$ for
the case where there exists $1.1\,\mathrm{TeV}$ Higgsino.\footnote
{
The expected significance is $3.5\sigma$ for $1.1\,\mathrm{TeV}$ Higgsino in our estimation.
Even though it is slightly below the $5\sigma$ discovery, we take $1.1\,\mathrm{TeV}$
Higgsino as an example because it is a candidate of the thermal relic DM.}

In Fig.~\ref{fig_c1_c2}, we show the contours of $1\sigma$ (dotted)
and $2\sigma$ (solid) constraints, which correspond to the values
$\sqrt{q}=1.9$ and $\sqrt{q}=2.8$, respectively, in the
$C_1~\mathrm{vs.}~C_2$ plane for $m=1.1\,\mathrm{TeV}$.  The blue,
green, and red lines denote the result obtained from the NC processes,
the CC processes, and the combined analysis, respectively.  The models
in the gray region are in more than $2\sigma$ tension with the
observation.  We also show several star markers that correspond to the
single $SU(2)_L$ multiplet contributions: the markers with ``$n_Y$''
represent an $SU(2)_L$ $n$-plet Dirac fermion with hypercharge $Y$,
while those with ``$n_\mathrm{Maj}$'' an $SU(2)_L$ $n$-plet Majorana
fermion.
Both the NC and CC constraints are represented
as straight bands in the $C_1~\mathrm{vs.}~C_2$ plane
since each process depends on a specific linear combination of $C_1$ and $C_2$.
In particular, the CC constraint is independent of $C_1$, or $Y$.
In this sense, the NC and CC processes are complementary to each other,
and thus we can separately constrain $C_1$ and $C_2$ only after
combining these two results.
For instance, we can exclude a single fermionic $SU(2)_L$ multiplet with $n \neq 2$ at more than
$2\sigma$ level, although
each process by itself cannot exclude the
possibility of $3_{\text{Maj}}$.  
We can also constrain the hypercharge, yet it is not uniquely determined.  
In addition to the Higgsino,
the EWIMP as an $SU(2)_L$ doublet Dirac fermion with
$|Y|^2\lesssim 2$ or an $SU(2)_L$ doublet Majorana fermion with
$|Y|^2\lesssim 5$ is still allowed.

\begin{figure}[t]
 \centering
 \includegraphics[width=0.48\linewidth]{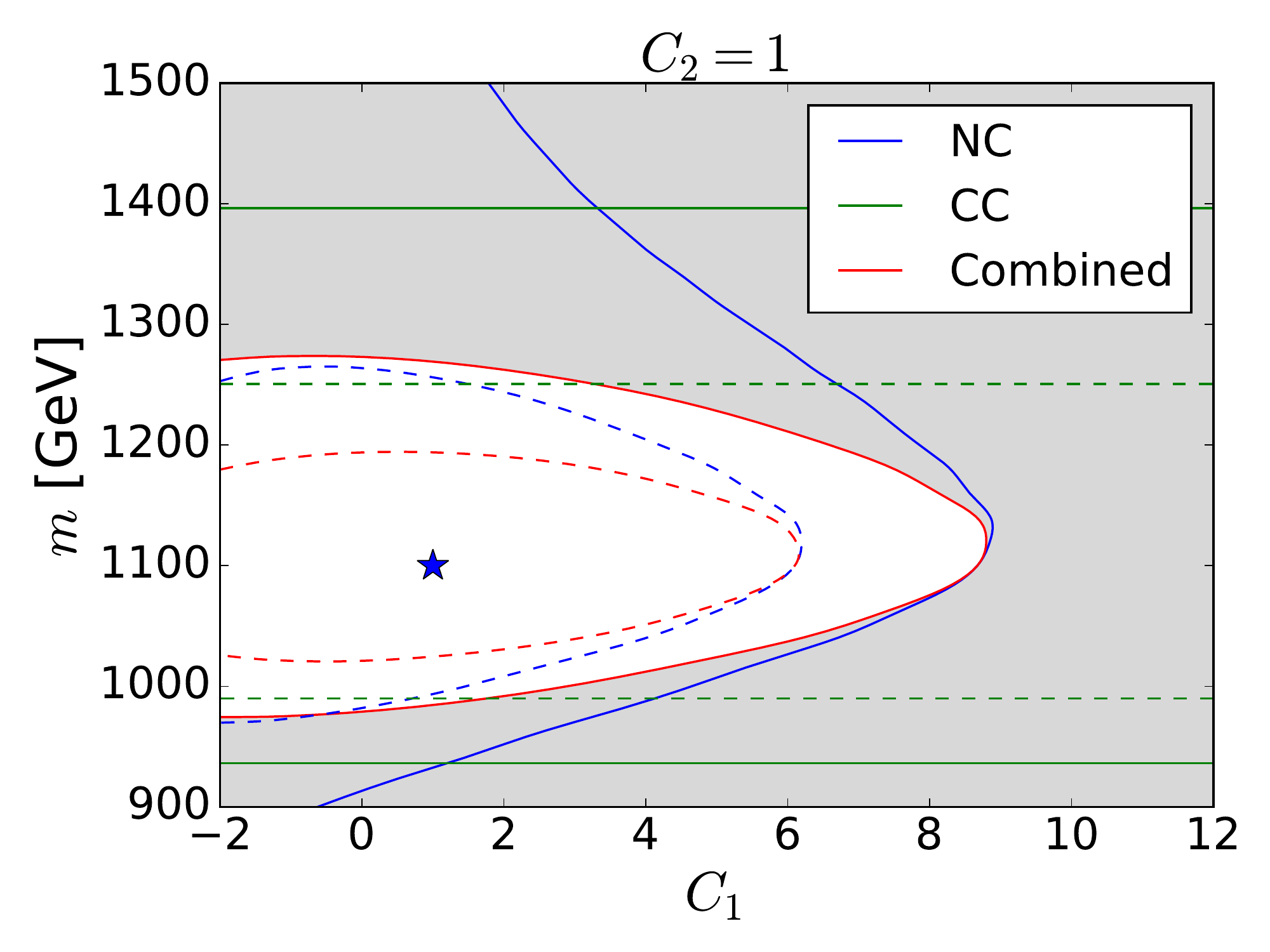}
 \includegraphics[width=0.48\linewidth]{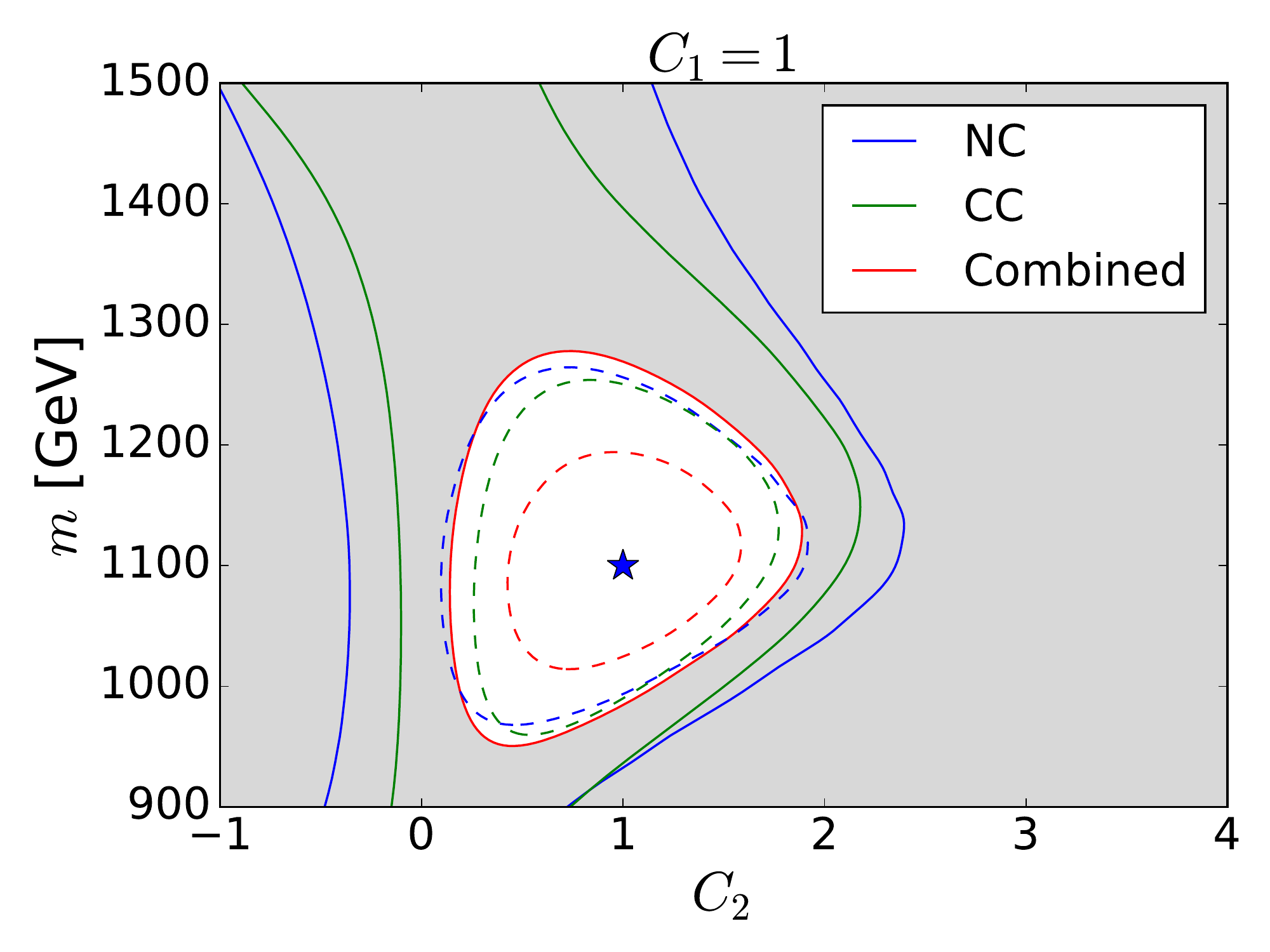}
 \caption{\textbf{Left:} Contour of $\sqrt{q}$ in the
 $C_1~\mathrm{vs.}~m$ plane with $C_2 = 1$,
 where we assume the $1.1\,\mathrm{TeV}$ Higgsino signal.
 The colors and
 styles of lines and the meaning of the gray region are the same as
 Fig.~\ref{fig_c1_c2}.  The star maker corresponds to the true Higgsino
 property $(C_1, m) = (1, 1.1\,\mathrm{TeV})$. \textbf{Right:} Contour
 of $\sqrt{q}$ in the $C_2~\mathrm{vs.}~m$ plane for $C_1 = 1$,
 where we assume the $1.1\,\mathrm{TeV}$ Higgsino signal.
The star maker corresponds to the true Higgsino property $(C_2, m) = (1,
1.1\,\mathrm{TeV})$.
 }
 \label{fig_c1_m}
\end{figure}

In Fig.~\ref{fig_c1_m}, we show the contour
plots of $\sqrt{q}$ in the $C_1~\mathrm{vs.}~m$ plane with $C_2=1$ (left)
and those in the $C_2~\mathrm{vs.}~m$ plane with $C_1=1$ (right).
The star marker in each panel shows the true values of
parameters $(C_1, m) = (1, 1.1\,\mathrm{TeV})$ (left) and $(C_2, m) =
(1, 1.1\,\mathrm{TeV})$ (right).  Again, by combining the NC and CC
results, we can significantly improve the determination of EWIMP
properties, making $1\sigma$ and $2\sigma$ contours closed circles in
the planes of our concern.  In particular, as red lines show, the
combined analysis allows us to determine the observed EWIMP mass at the
level of $\mathcal{O}(10)\%$.

\begin{figure}[t]
 \centering
 \includegraphics[width=0.5\linewidth]{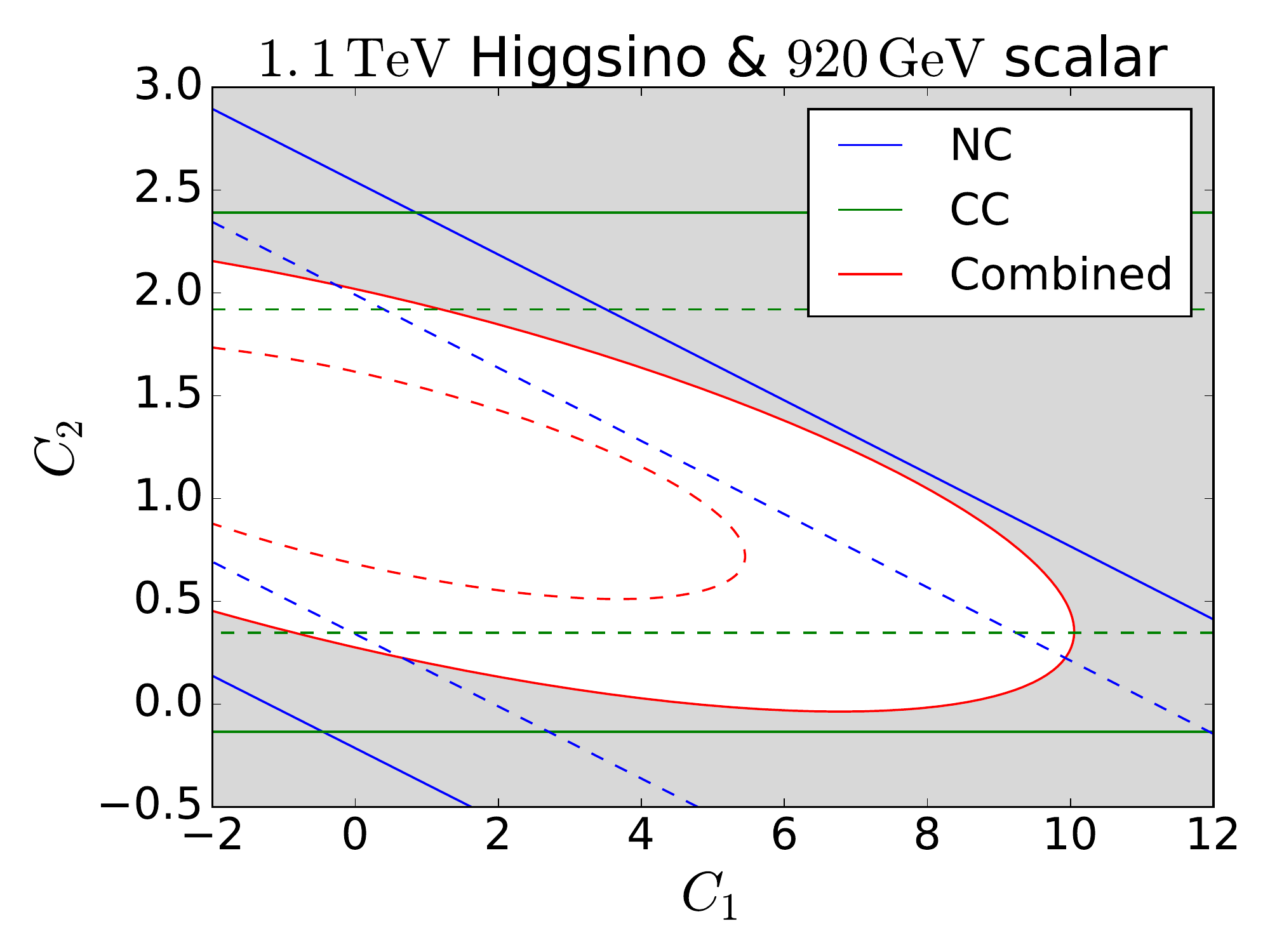}
 \caption{Contour of $\sqrt{q}$ in the $C_1~\mathrm{vs.}~C_2$ plane for
 the $1.1\,\mathrm{TeV}$ Higgsino signal, tested with the scalar EWIMP
 assumption.  The plane is defined as the scalar mass of $920\,\mathrm{GeV}$.
 The colors and styles of lines and the meaning of the gray region are
 the same as Fig.~\ref{fig_c1_c2}.}  \label{fig_c1_c2_scalar}
\end{figure}

\begin{figure}[t]
 \centering
 \includegraphics[width=0.48\linewidth]{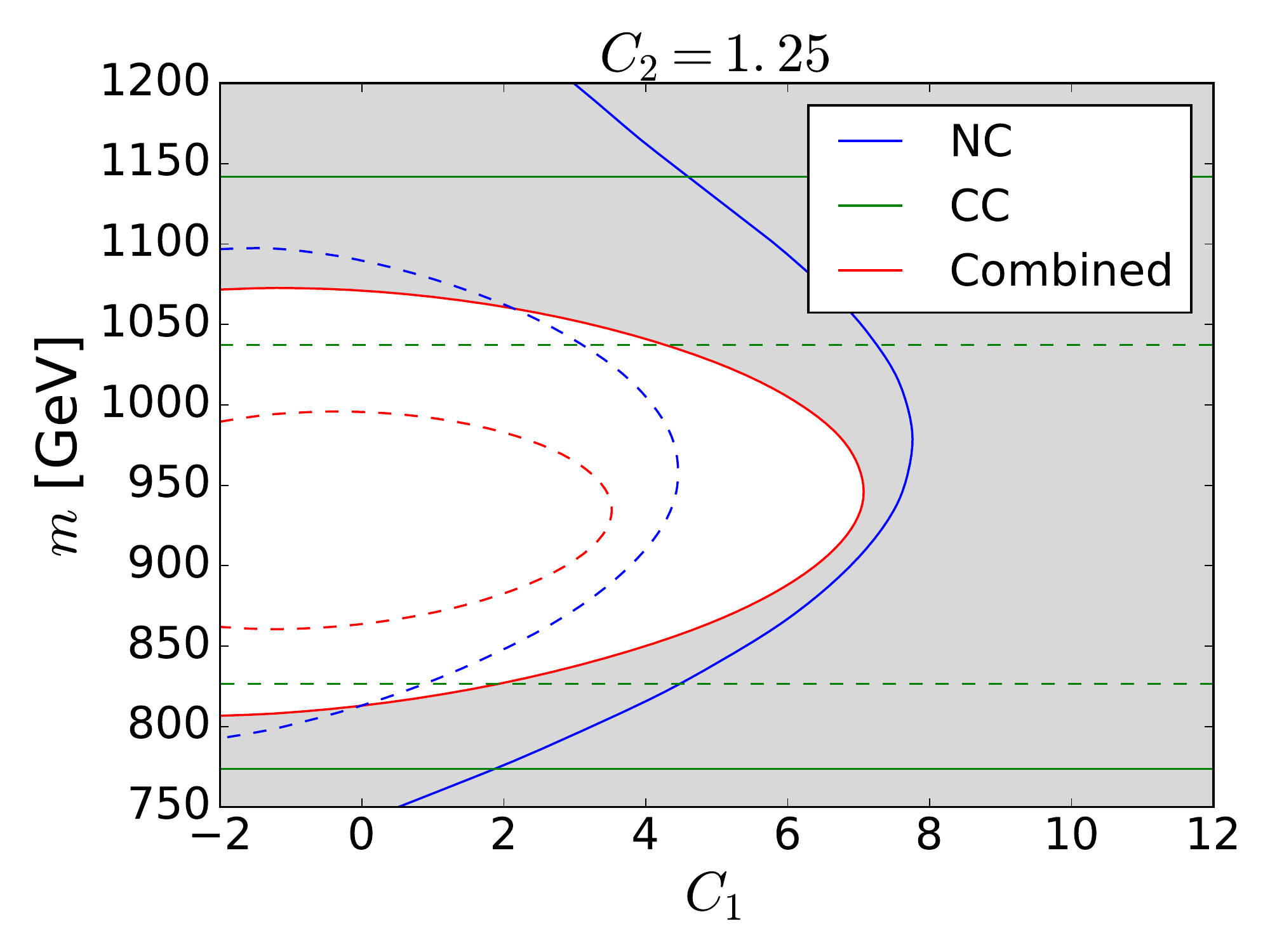}
 \includegraphics[width=0.48\linewidth]{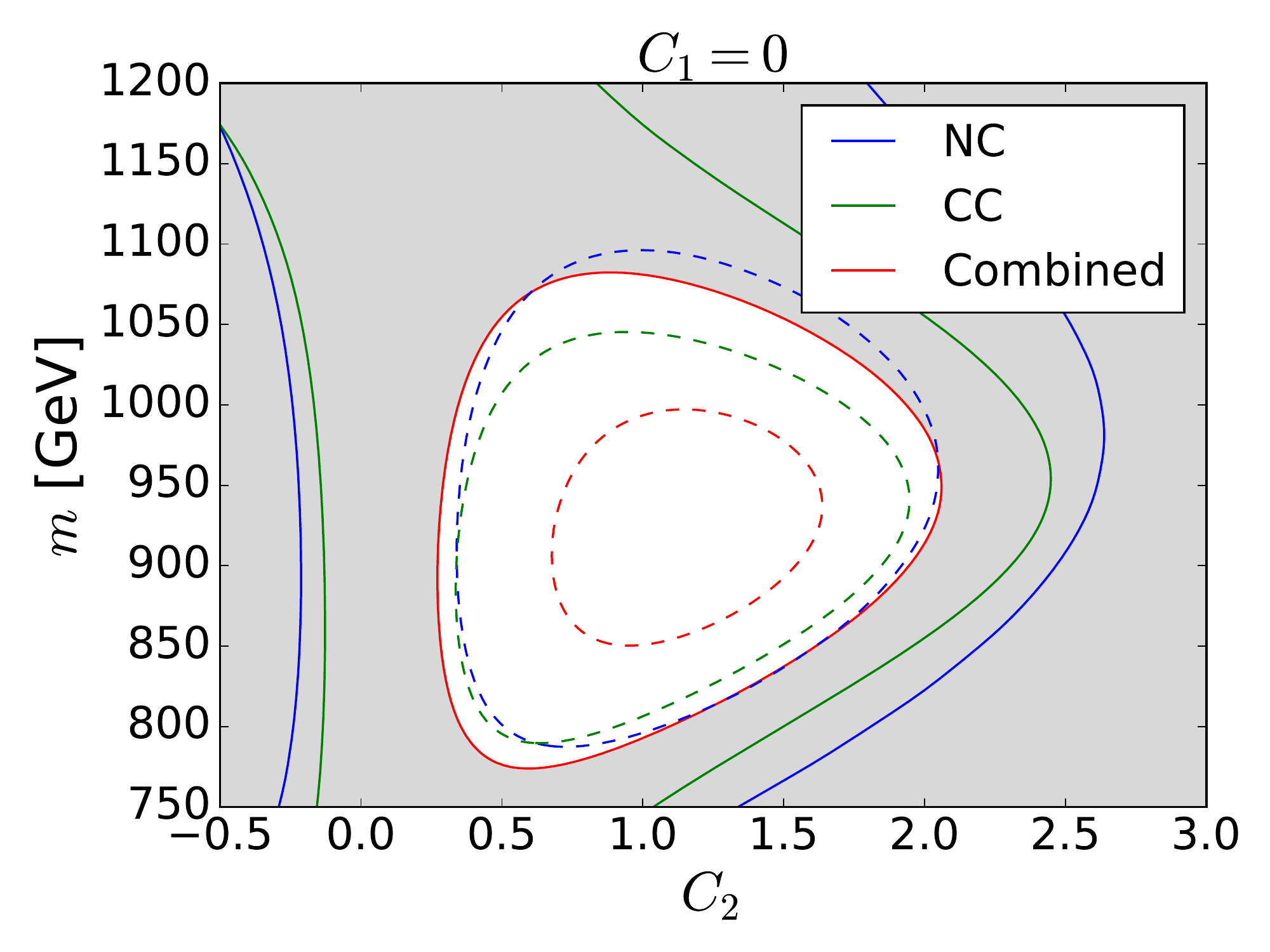}
 \caption{\textbf{Left:} Contour of $\sqrt{q}$ in the
 $C_1~\mathrm{vs.}~m$ plane with $C_2 = 1.25$ for the $1.1\,\mathrm{TeV}$ Higgsino signal,
 tested with the scalar EWIMP assumption.
 The colors and styles of lines and the meaning of the gray
 region are the same as Fig.~\ref{fig_c1_c2}.  \textbf{Right:} Contour
 of $\sqrt{q}$ in the $C_2~\mathrm{vs.}~m$ plane with $C_1 = 0$ for the
 $1.1\,\mathrm{TeV}$ Higgsino signal, tested with the scalar EWIMP
 assumption.} \label{fig_c1_m_scalar}
\end{figure}

Finally, we comment on the possibility of discriminating between
fermionic and scalar EWIMPs, whose difference comes from the loop
function $f(x)$ (see Eq.~\eqref{eq_f}).  Here we repeat the same
analysis explained above, assuming the $1.1\,\mathrm{TeV}$ Higgsino
signal for example, but use the scalar loop function to evaluate the
theoretical predictions $\bm{x}_f (m, C_1, C_2)$.  In
Figs.~\ref{fig_c1_c2_scalar} and \ref{fig_c1_m_scalar}, we show the
results in the $C_1$ vs. $C_2$ plane and the $C_1$ (or $C_2$) vs. $m$
plane, respectively, where one of the three parameters is fixed to its
best fit value.  It is seen that, in the case of the
$1.1\,\mathrm{TeV}$ Higgsino signal, it is hard to distinguish between
the bosonic and fermionic EWIMPs only with our method.  However, if a
part of the EWIMP properties (in particular its mass) is determined
from another approach, our method may allow us to determine its spin
correctly.

We also stress here that, with some favorable assumption about the
observed signal, we may obtain some hint about its spin.  For example,
if we assume that the observed signal composes a fraction of the dark
matter in our Universe, the choice of the EWIMP charges is
significantly constrained.  Note from Fig.~\ref{fig_c1_c2_scalar} that
the only choices of EWIMP charges that allow the EWIMP multiplet to
contain an electrically neutral component are
$(n,|Y|)=(3,0),(3,1),(4,1/2),(4,3/2)$, and $(5,0)_\text{real}$.  The
last column of the table~\ref{tab:minimalDM-for-950scalar-section} shows proper choices of EWIMP masses in order
for their thermal relic abundances become comparable with the dark
matter abundance in the current Universe.  All of those values are
somewhat larger than the central value of the mass of the observed
signal, which means that the scalar interpretation of the signal
cannot explain the whole of the dark matter relic abundance without
introducing some non-thermal production mechanism. 

\begin{table}[t]
\centering
\begin{tabular}{|c|ccc|}
\hline
$(n, Y)$           & $C_1$ & $C_2$ & $m_\text{DM}$[TeV] \\ \hline\hline
$(3,0)_\text{real}$  &    0  &  0.25  & 2.5 \cite{Farina:2013mla}             \\ 
$(3,          0)$  &    0  & 0.5   & 1.55 \cite{DelNobile:2015bqo}              \\ 
$(3,          1)$  & 0.75  & 0.5   & 1.6 \cite{Farina:2013mla}               \\ 
$(4,\frac{1}{2})$  & 0.25  & 1.25  & 2.4 \cite{Farina:2013mla}             \\ 
$(4,\frac{3}{2})$  & 2.25  & 1.25  & 2.9 \cite{Farina:2013mla}             \\ 
$(5,0)_\text{real}$  &    0  &  1.25  & 9.4 \cite{Farina:2013mla}          \\ 
\hline
\end{tabular}
\caption{The scalar EWIMPs that are compatible with the result in Fig.~\ref{fig_c1_c2_scalar}. The observed DM energy density is explained by the
thermal relic of the EWIMP with $m_{\text{DM}}$ shown in the fourth column.}
\label{tab:minimalDM-for-950scalar-section}
\end{table}

\section{Conclusion}
\label{seq:conclusion}

In this paper, we have discussed the indirect search of EWIMPs at future
$100\,\mathrm{TeV}$ hadron colliders based on the precision measurement
of the production processes of a charged lepton pair and that of a
charged lepton and a neutrino.  In particular, we have demonstrated that
not only we can discover the EWIMPs, but also we can determine their
properties such as their masses, $SU(2)_L$ and $U(1)_Y$ charges, and
spins via the processes of our concern.  It is based on two facts: the
high energy lepton production channel enables us to study its momentum
distribution in great detail, and the EWIMP correction shows
characteristic features, including a dip-like structure as the final
state invariant mass being twice the EWIMP mass.  The latter feature
also helps us to distinguish the EWIMP signals from backgrounds and
systematic errors, as they are not expected to show a dip-like
structure.  In order to fully exploit the differences between the
distributions the EWIMP signals and systematic errors, we have adopted
the fitting based analysis as our statistical treatment.

First, we have shown in Fig.~\ref{fig_mchi_vs_sqq0} the detection
reach of Higgsino and Wino from the neutral current (NC) processes
(mediated by photon or $Z$-boson), the charged current (CC) processes
(mediated by $W$-boson), and the combination of these two results.  We
have seen that the addition of the CC processes improves the detection
reach from the previous analysis \cite{Chigusa:2018vxz}.  From the
combined analysis, the bounds at the $5\sigma$ ($95\%$ C.L.) level for
Higgsino and Wino are $850\,\mathrm{GeV}$ ($1.7\,\mathrm{TeV}$) and
$1.3\,\mathrm{TeV}$ ($2.3\,\mathrm{TeV}$), respectively.  This result,
in particular that for short lifetime Higgsino, indicates the
importance of our method for the EWIMP search.

Next, we have considered the determination of the mass and $SU(2)_L$
and $U(1)_Y$ charges of the observed EWIMP.  By combining the NC and
the CC events, the position and the height of the dip in the EWIMP
effect on the cross section gives us enough information for
determining all the three parameters.  In Figs.~\ref{fig_c1_c2} and
\ref{fig_c1_m}, we have shown the plots of the test statistics that
test the validity of several choices of parameters.  As a result, the
$SU(2)_L$ charge of the observed signal is correctly identified under
the assumption of a single EWIMP multiplet, and the $U(1)_Y$ charge
and mass are also determined precisely.  In order for the
determination of the EWIMP spin, we have plotted the contours of the
test statistics that test the validity of the scalar EWIMP models with
some fixed values of masses and charges.  The results are shown in
Figs.~\ref{fig_c1_c2_scalar} and \ref{fig_c1_m_scalar}, which reveals
that the spin is not completely determined by solely using our
method.  Use of another approach to determine the EWIMP properties, or
of some assumption like that the observed signal corresponds to the
dark matter in our Universe, may help us to obtain further information
regarding the EWIMP spin.

\section*{Acknowledgments}

This work was supported by JSPS KAKENHI Grant (Nos.\ 16K17715 [TA], 17J00813 [SC],
16H06490 [TM], and 18K03608 [TM]).

\bibliographystyle{elsarticle-num}
\bibliography{fccwimp}

\end{document}